\documentclass[conference]{IEEEtran}
\IEEEoverridecommandlockouts

\usepackage{cite}
\usepackage{amsmath,amssymb,amsfonts}
\usepackage{algorithmic}
\usepackage{graphicx}
\usepackage{subfigure}
\usepackage{textcomp}
\usepackage{xcolor}
\usepackage[a4paper, total={184mm,239mm}]{geometry}
\usepackage{booktabs}
\usepackage{multirow}
\def\BibTeX{{\rm B\kern-.05em{\sc i\kern-.025em b}\kern-.08em
    T\kern-.1667em\lower.7ex\hbox{E}\kern-.125emX}}

\usepackage[small,compact]{titlesec} 
\usepackage[compact]{titlesec}
\titlespacing{\section}{2pt}{2pt}{2pt}
\titlespacing{\subsection}{2pt}{2pt}{2pt}
\titlespacing{\subsubsection}{2pt}{2pt}{2pt}

\begin{document}

\title{OplixNet: Towards Area-Efficient \underline{Op}tical\\Sp\underline{li}t-Comple\underline{x} Networks with Real-to-Complex\\Data Assignment and Knowledge Distillation 
}

\author{
\IEEEauthorblockN{Ruidi Qiu$^1$,
Amro Eldebiky$^1$,
Grace Li Zhang$^2$,
Xunzhao Yin$^3$,
Cheng Zhuo$^3$,
Ulf Schlichtmann$^1$,
Bing Li$^1$}
\IEEEauthorblockA{$^1$Technical University of Munich, $^2$TU Darmstadt, $^3$Zhejiang University}
\IEEEauthorblockA{Email: \{r.qiu, amro.eldebiky, ulf.schlichtmann, b.li\}@tum.de, grace.zhang@tu-darmstadt.de, \{xzyin1, czhuo\}@zju.edu.cn}
}


\maketitle
\begin{abstract}
Having the potential for high speed, high throughput, and low energy cost, optical neural networks (ONNs) have emerged as a promising candidate for accelerating deep learning tasks. In conventional ONNs, light amplitudes are modulated at the input and detected at the output. However, the light phases are still ignored in conventional structures, although they can also carry information for computing. To address this issue, in this paper, we propose a framework called OplixNet to compress the areas of ONNs by modulating input image data into the amplitudes and phase parts of light signals.
The input and output parts of the ONNs are redesigned to make full use of both amplitude and phase information. Moreover, mutual learning across different ONN structures is introduced to maintain the accuracy. Experimental results demonstrate that the proposed framework significantly reduces the areas of ONNs with the accuracy within an acceptable range. For instance, 75.03\% area is reduced with a 0.33\% accuracy decrease on fully connected neural network (FCNN) and 74.88\% area is reduced with a 2.38\% accuracy decrease on ResNet-32. 
\end{abstract}


\section{Introduction}
\label{intro}

Artificial neural networks (ANN) have garnered significant achievements across numerous domains, including computer vision \cite{lenet} and natural language processing \cite{lstm}. 
Recently, the development of emerging large language models, such as ChatGPT, caused a higher demand for neuromorphic computing devices. 
Besides CPUs, GPUs, TPUs \cite{tpu} and other ASICs, some emerging devices, such as superconducting circuits \cite{superconductingAI}, RRAM \cite{YiyuDAC2022,AICAS2020,Correct2023,Yiyu2022,Yiyu2023}, and optical devices \cite{ONN_the_first, zhu2020,  amro2023, onn_imprecision} are also proposed to further boost computing speed, reduce energy consumption and save device area of ANN computations.

Optical neural network (ONN) is a new neuromorphic computing solution based on optical components. The ONN structure proposed in \cite{ONN_the_first} uses arrays of Mach-Zehnder Interferometers (MZIs) to perform the matrix-vector multiplications (MVMs). The weight matrices obtained from software-trained networks are mapped to the phases of the MZIs via singular value decomposition (SVD) and unitary-to-interferometer mapping \cite{onn_unitary_mapping_triangle}. The amplitudes and phases of the light signal are changed while going through an MZI array to implement multiplications and additions. These computations can be conducted in an extremely low latency with an optical detection rate over 100 GHz \cite{onn_100Ghz}.

However, the previous ONN structure suffers from large area and static power consumption. To implement an m$\times$n weight matrix, the conventional MZI-ONN \cite{ONN_the_first} requires 
$\frac{n(n-1)}{2}+min(m,n)+\frac{m(m-1)}{2}$ MZIs. 
In addition, the power consumption is highly related to the number of MZI devices. To maintain the phase value, a high static power is consumed in each MZI ($0\sim80$ mW per PS depending on the phase value \cite{complex_onn}). To improve the efficiency of ONNs, various approaches and structures have been proposed, including replacing traditional optical U$\Sigma$V structure \cite{onn_slim}, pruning \cite{onn_pruning_LTH, onn_OFFT}, and employing FFT structures \cite{onn_OFFT}. However, the above approaches cannot have a stable area reduction ratio without significant accuracy degradation. 
For example, the average reduction ratio of optical devices (including directional couplers and phase shifters) in \cite{onn_OFFT} is 39.16\%, and the lowest reduction ratio is only 19.69\%. 
The pruning approach proposed in \cite{onn_pruning_LTH} obtains a sparsity of 88.89\% but causes a 4.99\% accuracy degradation, which is too large for fully-connected neural networks (FCNN). 

As the light pulses in ONN contain both amplitude and phase information, all the values in ONN can be represented in complex form. 
However, most of the previous work only considers the amplitude at the input modulator, leaving the phase of the input signal unchanged. 
By assigning real data onto amplitudes and phases, the number of input light signals can be decreased, so that the area of the ONN can be reduced. 
This complex ONN idea is mentioned in \cite{complex_onn, onn_imprecision}. 
However, both of them only consider the simplest models, such as FCNNs and XOR classifiers, and ignore the data assigning schemes. 

In this paper, a framework is presented to exploit the potential of complex encoding of the input data in ONNs to reduce feature sizes. The framework considers the trade-off in various aspects, including the area occupation, the inference accuracy and the implementation of optical hardware. The contributions of this paper are summarized as follows:
\begin{itemize}
    \item 
    A method to assign real data to the amplitudes and phases of light signals based on different spatial or channel arrangements is explored. 
    Our data assigning scheme realizes around 75\% area reduction with minimal accuracy degradation by reducing the size of the feature map. 
    \item 
    A directional-coupler-based optical complex encoder is implemented to encode input data into amplitudes and phases of light signals. Compared with previous work, the proposed encoder is able to handle high throughput data.
    \item
    A knowledge transfer approach is presented for mutual learning between split complex-value neural network (SCVNN) with complex input assignment and the conventional complex-valued neural network (CVNN) without complex input assignment to maintain the inference accuracy.
    \item
    An optical learnable complex-to-real decoder is proposed to ease the detection of the complex output signal. Unlike the previous fixed detection approach, our decoder does not require any further processing and achieves higher accuracy with only 0.04\%$\sim$0.73\% more area occupation.
\end{itemize}

The rest of the paper is organized as follows. In Section \ref{sec:background}, the background is explained. 
In Section \ref{sec:method}, the proposed framework considering area, performance and hardware implementation is elaborated. 
Experimental results are shown in Section \ref{sec:experiments} and conclusions are drawn in Section \ref{sec:conclusion}.

\section{Preliminaries}
\label{sec:background}
\subsection{Structure of MZI-based ONN}
\label{subsec:mzi-onn}
\begin{figure}
    \centering
    \subfigure[]{
        \includegraphics[scale=1.45]{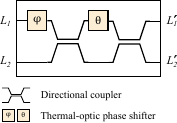}
        \label{fig:MZI_structure}
    }
    \subfigure[]{
        \includegraphics[scale=1.45]{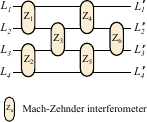}
        \label{fig:MZI_array}
    }
    \quad
    \subfigure[]{
        \includegraphics[scale=1.4]{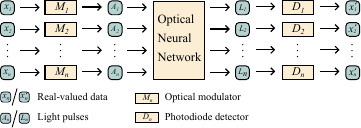}
        \label{fig:MZI_ONN_conventional}
    }
    \vspace{-2mm}
    \caption{Structures of ONN components and conventional ONN data flow: (a) MZI structure; (b) MZI array structure; (c) data flow in conventional ONN;}
    \vspace{-4mm}
    \label{fig:structures_MZI_MZIArray_ONN}
\end{figure}
The MZI array is the basic component of ONN. An MZI is composed of two 50:50 directional couplers (DC) and two thermal-optic phase shifters (PS) \cite{ONN_the_first}, as shown in Figure \ref{fig:MZI_structure}. 
The DC transmits half of the energy from input light pulses to each output port and adds a $\frac{\pi}{2}$ phase shift to the diagonal transmission.
The PS is tuned by integrated heaters to conduct a certain phase shift on light signals \cite{complex_onn}. 
Thus, an MZI can implement a 2$\times2$ unitary matrix by tuning PSs and conduct MVM, which is:
\begin{equation}
\hspace{-1.9mm}
\begin{bmatrix}
    L^{\prime}_1 \\
    L^{\prime}_2
\end{bmatrix}
=
\begin{bmatrix}
    \frac{1}{\sqrt{2}} & \frac{i}{\sqrt{2}} \\
    \frac{i}{\sqrt{2}} & \frac{1}{\sqrt{2}}
\end{bmatrix}
\begin{bmatrix}
    e^{i\theta} & 0 \\
    0 & 1
\end{bmatrix}
\begin{bmatrix}
    \frac{1}{\sqrt{2}} & \frac{i}{\sqrt{2}} \\
    \frac{i}{\sqrt{2}} & \frac{1}{\sqrt{2}}
\end{bmatrix}
\begin{bmatrix}
    e^{i\phi} & 0 \\
    0 & 1
\end{bmatrix}
\begin{bmatrix}
    L_1 \\
    L_2
\end{bmatrix}
.
\end{equation}
According to \cite{onn_unitary_mapping_triangle, onn_unitary_mapping_square}, by tuning the phase shifts of PSs, an MZI array can be used to implement arbitrary \textit{unitary} matrix. 
Figure \ref{fig:MZI_array} illustrates the implementation of a $4\times4$ unitary matrix using 6 MZIs. A general weight matrix can be decomposed using SVD: $W = U\Sigma V^{*}$. $U$ and $V^{*}$ are unitary matrices, and $\Sigma$ is a diagonal matrix. Therefore, it can be implemented in ONN by using MZI arrays and optical attenuators. The number of MZIs required for an $m\times n$ matrix is the summation of MZI numbers of $U$, $\Sigma$ and $V^{*}$, which is $\frac{n(n-1)}{2}+min(m,n)+\frac{m(m-1)}{2}$. 

\subsection{Conventional ONN and complex-valued ONN}
\label{subsec:complex onn}

In this paper, various types of neural networks are discussed. To make the elaboration more clear, our definitions of these NNs are shown in Table \ref{tab:definitions_ONN_CVNN_SCVNN_RVNN}. 
Note that CVNN is the software model that can be deployed onto conventional ONN by SVD and unitary-to-interferometer mapping, and SCVNN can be deployed onto complex ONN in the same way.

\begin{table}[t]\scriptsize
    \vspace{-2mm}
    \caption{Definitions of neural networks related in this paper}
    \begin{center}
    \begin{tabular}{cc}
        \toprule
        Definitions & Descriptions \\
        \toprule
        \multirow{2}*{Conventional ONN} & the ONNs only encoding the input data\\& into amplitudes of light signals, such as \cite{ONN_the_first};\\
        \cmidrule{1-2}
        \multirow{2}*{Split ONN} & the ONNs encoding both amplitudes and phases\\& of light signals, such as our proposed framework;\\
        \cmidrule{1-2}
        \multirow{2}*{CVNN} & the software-trained complex-valued neural networks;\\& only encoding the real parts of complex input values;\\
        \cmidrule{1-2}
        \multirow{2}*{SCVNN} & the software-trained split complex-valued neural networks;\\& encoding two real input values into one complex value;\\
        \cmidrule{1-2}
        RVNN & the software-trained real-valued neural networks;\\
        \bottomrule
    \end{tabular}
    \end{center}
    \label{tab:definitions_ONN_CVNN_SCVNN_RVNN}
    \vspace{-5mm}
\end{table}

The input encoders of the conventional ONNs, such as \cite{ONN_the_first}, only modulate the amplitude of light signals, while leaving the light signals with the same phase values, as is shown in Figure \ref{fig:encoders}(c). 
The input data from the dataset are assigned to the amplitude of light signals. 
At the output part of the conventional ONNs, photodiodes are used as the decoders to detect the amplitudes of output light signals but discard the phase information.
The data flow of the conventional ONN is shown in Figure \ref{fig:MZI_ONN_conventional}.

To fully utilize the complex property of ONNs, an idea of complex ONN is proposed in \cite{complex_onn}. At the input of the complex ONN, DCs and PSs are used to encode both amplitudes and phases of light signals. 
However, the appearance of PSs means that the temperature must be changed frequently when a large amount of data is input, which becomes a bottleneck when handling high throughput data. 

To assign real data to the complex input data, a naive input data assigning method is mentioned in \cite{onn_imprecision}. 
In each channel, the input image data is spatially split into top and bottom parts. 
Two values from the same position in the top and bottom parts are assigned to the real and imaginary parts of a complex value. For example, the top left points from the top and bottom parts are assigned together, as is shown in Figure \ref{fig:spatial data assignment}(b).

To detect the complex information of output light signals, a coherent detection approach is proposed in \cite{complex_onn} where a reference light signal with known amplitude and phase is given to interfere with the output optical signals, as is depicted in Figure \ref{fig:decoders}(c). 
However, this approach needs additional time to conduct a phase shift on the reference light. In addition, post-processing is needed after photodiode detection. 
The same problems also exist in the coherent phase-detection schemes in \cite{complex_onn_1994}.

\section{Methodology}
\label{sec:method}
\begin{figure}[t]
    \centering
    \includegraphics[scale=1]{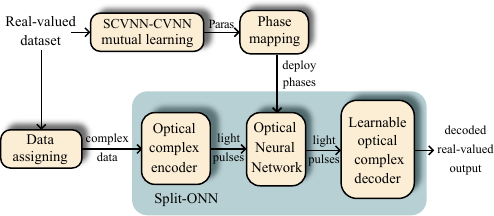}
    \caption{The workflow of OplixNet}
    \label{fig:workflow}
    \vspace{-0.5cm}
\end{figure}

In this section, the basic idea of split ONN is first discussed. 
The input encoder of the proposed framework is then introduced.
Due to fewer parameters used, the accuracy of our framework undoubtedly decreases. To restore the inference accuracy, a knowledge distillation method between CVNN and SCVNN is adopted. 
At last, a learnable complex optical output decoders is proposed. 
The workflow of the proposed method is shown in Figure \ref{fig:workflow}.

\subsection{Optical split complex-valued neural network}
\label{subsec: scvnn}
In this work, we consider not only the amplitudes but also the phases of the light signals when encoding input data at the input of an ONN. 
Although the original dataset is real-valued, we assign the real values to the real and imaginary parts of the complex data. 
In this way, our ONN becomes split ONN and is equivalent to SCVNN \cite{cvnn_survey} on software. 
A $2\times2$ complex-valued matrix $\mathbf{W_c}$ in SCVNN can be converted to a $4\times4$ real-valued matrix $\mathbf{W_{cr}}$. 
Therefore, the complex MVM can be converted to the real MVM with more dimensions. 
A template complex-to-real MVM conversion is shown below:
{
\setlength\abovedisplayskip{0.25cm}
\setlength\belowdisplayskip{0.25cm}
\begin{equation}\small
\begin{split}
    \mathbf{W_c}\mathbf{x_c} 
    \!&=
    \begin{bmatrix}
        w_{11}\!+\!jw_{12} & w_{13}\!+\!jw_{14} \\
        w_{21}\!+\!jw_{22} & w_{23}\!+\!jw_{24}
    \end{bmatrix}
    \!
    \begin{bmatrix}
        x_1\!+\!jx_2 \\
        x_3\!+\!jx_4
    \end{bmatrix}
    \\
    &=
    \begin{bmatrix}
        x_1^{\prime}+jx_2^{\prime} \\
        x_3^{\prime}+jx_4^{\prime}
    \end{bmatrix}
    =
    \mathbf{x_{c}^{\prime}}
    \\
    \rightarrow
    \mathbf{W_{cr}}\mathbf{x_{cr}} 
    \!&= 
    \setlength\arraycolsep{2pt}
    \!
    \begin{bmatrix}
        w_{11} & -w_{12} & w_{13} & -w_{14} \\
        w_{12} & w_{11}  & w_{14} & w_{13}  \\
        w_{21} & -w_{22} & w_{23} & -w_{24} \\
        w_{22} & w_{21}  & w_{24} & w_{23}
    \end{bmatrix}
    \!
    \begin{bmatrix}
        x_1 \\ x_2 \\ x_3 \\ x_4
    \end{bmatrix}
    \!= \!
    \begin{bmatrix}
        x_{1}^{\prime} \\ x_{2}^{\prime} \\ x_{3}^{\prime} \\ x_{4}^{\prime}
    \end{bmatrix}
    \!=\! \mathbf{x_{cr}^{\prime}}
    ,
\end{split}
\label{for:matrices_scvnn}
\end{equation}
}
where $\mathbf{x_c}$ and $\mathbf{x_{cr}}$ denote the input vectors in real and complex form, respectively. $\mathbf{x_c^{\prime}}$ and $\mathbf{x_{cr}^{\prime}}$ denote the corresponding output vectors.
Although having the same number of elements, the $4\times4$ matrix $\mathbf{W_{cr}}$ has less expressiveness than a normal real-valued $4\times4$ matrix in RVNN. 
This is because $\mathbf{W_{cr}}$ only has half the number of independent weights compared with normal $4\times4$ in RVNN, which has 16 independent weights. 
This difference is from the relevance of the real and imaginary parts of complex data.
Taking $w_{11}$ as an example, it has an impact on both $x_1^{\prime}$ and $x_2^{\prime}$, which are the real and imaginary parts of one complex value.
This relevance boosts the network performance if the original dataset is naturally in complex form, such as wireless communication or audio processing \cite{cvnn_compare_with_rvnn, cvnn_survey}.
However, if two real values without prior relevance are assigned to one complex value, then the additional relevance from the real and imaginary parts will cause an accuracy degradation. 
Therefore, an assigning method that assigns two real values with as much relevance as possible to one complex value is important to split ONN.

\begin{figure}[t]
    \centering
    \includegraphics[scale=0.95]{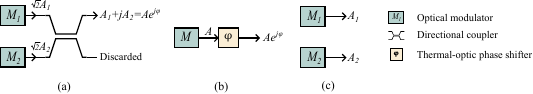}
    \caption{Implementation of encoders: (a) DC-based complex encoder (proposed); (b) PS-based complex encoder; (c) real-valued encoder in conventional ONN;}
    \vspace{-0.1cm}
    \label{fig:encoders}
\end{figure}

\subsection{Optical input encoder}
\label{subsec: optical input encoder}
\subsubsection{Hardware implementation of input encoder}

Our optical complex encoder aims to encode input data into both amplitude and phase parts.
The naive way is to use modulators to modulate amplitudes and one PSs to conduct phase shifts, as is shown in Figure \ref{fig:encoders}(b). 
However, this encoder has a time bottleneck problem on PSs, which is the same as \cite{complex_onn} and has already been discussed in Section \ref{subsec:complex onn}.  
As shown in Figure \ref{fig:encoders}(a), a DC-based optical complex encoder using no thermal-optic PS is proposed to overcome the PS time bottleneck.
The two input light signals have the same static phase and are modulated by two separate modulators, respectively. The phase of light is a relative value, and all the phase values of light signals mentioned in ONN are relative to the initial static phase. Thus, the original phase of input signals can be defined as 0. Then, the optical signals after modulation can be represented as $\sqrt{2}A_1$ and $\sqrt{2}A_1$.
The power dividing ratio of the DC is set to 50:50, and the phase of the bottom input data is shifted by 90 degrees. Therefore, the output at the top port is $A_1 + jA_2$,
which contains the information from two real input signals but has no time bottleneck when dealing with high throughput input data.

\begin{figure}[t]
    \centering
    \includegraphics[scale=1.35]{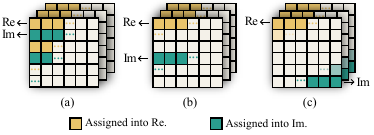}
    \vspace{-0.3cm}
    \caption{Spatial data assignment (taking three-channel image data as an example): (a) spatial interlace (proposed); (b) spatial half-half; (c) spatial symmertric;}
    \label{fig:spatial data assignment}
    \vspace{-0.4cm}
\end{figure}

\subsubsection{Spatial data assignment for FCNNs}
To apply real-valued datasets to our split ONN, real-to-complex data assignment before hardware encoding is needed. 
For FCNN-MNIST task, a \textit{spatial interlace} data assigning method is proposed.
As is depicted in Figure \ref{fig:spatial data assignment}(a), in each channel of the original image, every two adjacent real points (top and bottom) are assigned to a complex point. 
The channel information of these complex points remains consistent with the original. 
As is discussed in Section \ref{subsec: scvnn}, the more relevance that two values have, the less accuracy degradation they cause after being assigned to one complex value. Therefore, assigning two neighboring points from the original image is better than assigning two points from the top half and bottom half. 
This spatial assigning method compresses the height of the image spatially by half. 
In this way, the number of input values and the size of mid-layer feature maps become half. 
Therefore, the size of the linear layer in FCNN will decrease from  $m\times n$ to  $\frac{m}{2}\times \frac{n}{2}$ and the total size of our split ONN reduces almost $75\%$ compared with the conventional ONN.
To demonstrate the impact of distance before assigning, another spatial assigning method, called \textit{spatial symmetric}, is compared with our spatial interlace method. As shown in Figure \ref{fig:spatial data assignment}(c), the points in the upper left corner and the lower right corner are assigned to the same complex point. The area reduction of this method is the same as spatial interlace. The experimental performance difference of these spatial data assigning methods is demonstrated in Section \ref{sec:experiments}.

However, the spatial assigning methods cannot decrease the area of CONV layers by compressing the size of feature maps. The size of CONV kernel is only related to the number of input and output channels and the spatial size of CONV kernel, such as $3\times 3$ or $5\times 5$. Thus, data assignment from other dimensions of the input data is needed.


\begin{figure}[t]
    \centering
    \includegraphics[scale=1.05]{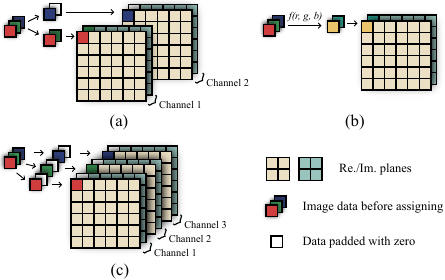}
    \vspace{-0.3cm}
    \caption{Channel data assignment: (a) channel lossless (proposed); (b) channel remapping; (c) conventional ONN;}
    \label{fig:channel data assignment}
    \vspace{-0.5cm}
\end{figure}

\subsubsection{Channel data assignment for CNNs}
For conventional CNN models, such as LeNet-5 and ResNet series, a channel assigning method \textit{channel lossless} is proposed to assign points at the same spatial position but in different channels into one complex point. As is shown in Figure \ref{fig:channel data assignment}(a), for the input image data with three color channels, the first two color channels are assigned to the real and imaginary parts of the first complex channel. The third color channel is assigned into the real part of the second complex channel, leaving the imaginary part padded with zeros. The data assigning method of conventional ONN is shown in Figure \ref{fig:channel data assignment}(c) to further indicate the difference between our method and the conventional method.
With the channel lossless assigning method, the channel number of the feature maps in mid-layers decreases to half compared with conventional ONN because each complex channel contains two real channels. In each CONV layer, the input and output channel number of the CONV kernel also becomes half. Thus, the channel lossless assigning method used in OplixNet achieves around 75\% area reduction ratio in terms of MZI numbers.

To explore more channel assigning methods, a \textit{channel remapping} method is proposed to compare with our channel lossless method. As is shown in Figure \ref{fig:channel data assignment}(b), the three color channels in the original image are first mapped into two real channels. Then the points in the first real channel are assigned to the real part of the complex channel, and the second channel goes to the imaginary part. The $f(r, g, b)$ is the mapping function from three channels to two channels. This assigning method generates a thinner feature map than our channel lossless method but causes severe accuracy degradation because some information is lost during the lossy mapping. 

\subsection{SCVNN-CVNN mutual learning}

As discussed in Section \ref{subsec: optical input encoder}, no matter which data assigning approach is used, there is an accuracy degradation compared with the original optical CVNN. 
In this paper, knowledge distillation \cite{knowledge_distillation} is used to restore the accuracy of SCVNN. We choose the CVNN model as our teacher model. To make the teacher more knowledgeable, a larger model in the same series is chosen. For instance, the CVNN version of ResNet-56 is chosen as the teacher of our ResNet-32 SCVNN model.

To make the teacher better understand the student, we use mutual learning \cite{mutual_learning} to train both models. The teacher and student start from the beginning and learn knowledge from training data and the other model. The loss function of both models can be expressed as:
\begin{equation}
    L_{SCVNN} = L_{CE} + \alpha L_{KD\_CVNN},
\end{equation}
\begin{equation}
    L_{CVNN} = L_{CE} + \alpha L_{KD\_SCVNN},
\end{equation}
where $L_{CE}$ denotes the cross entropy loss from training data. $L_{KD\_CVNN}$ and $L_{KD\_SCVNN}$ denote the distillation loss from the other network. $\alpha$ is the mixing factor of the two losses.

\subsection{Optical output decoding}

\begin{figure}
    \centering
    \includegraphics[scale=1.25]{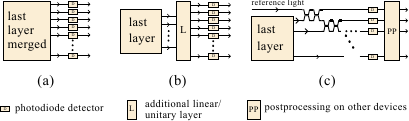}
    \caption{Learnable decoders and the coherent detection decoder: (a) learnable ``merging'' decoder (proposed); (b) learnable linear/unitary decoder; (c) coherent detection decoder;}
    \label{fig:decoders}
    \vspace{-5mm}
\end{figure}

The decoders in ONN aim to extract the information from output light pulses.
The previous optical complex decoders in Section \ref{subsec:complex onn} are all unlearnable. Due to the drawbacks of these decoders discussed before, learnable optical decoders are adopted in our ONN circuit. The learnable decoders are inserted into the network structure, and their parameters are learned during the training stage. The input of decoders is the light signals with amplitude and phase information. After decoding, the complex information is converted into amplitude information and can be detected by photodiodes. As is shown in Figure \ref{fig:decoders}(b), The most naive learnable decoder is to add an additional layer, such as a linear layer or unitary layer (can be implemented by an MZI array in ONN), to the end of the network. In our OplixNet framework, a learnable \textit{merging} decoder is proposed for all NN models. As shown in Figure \ref{fig:decoders}(a), this design merges the decoder layer into the last layer of the neural network. In the merged last layer, the output size will double, but the input size will stay the same. Compared with the linear/unitary decoder, it has more weight parameters. However, in optical circuits, it costs less MZIs than linear/unitary decoder, which means it has a better inference accuracy with lower area cost. The MZI cost can be computed using the formula mentioned in Section \ref{subsec:mzi-onn}



\section{Experimental Results}
\label{sec:experiments}
To evaluate the efficiency and performance of our OplixNet, four different NN models, an FCNN with a hidden layer size of 100, LeNet-5, and two deep residual networks, ResNet-20 and ResNet-32, are tested on different datasets, MNIST, CIFAR-10, and CIFAR-100. The neural networks were trained with NVIDIA A100 and A30 GPUs. To make the comparison fair, for each NN model, experiments with different settings are run with the same hyperparameters. 

\begin{table}[t]\scriptsize
    \caption{Experimental results of proposed work}
    \vspace{-0.4cm}
    \begin{center}
    \begin{tabular}{*{7}{c}}
        \toprule
        \multirow{2}*{Model} & \multicolumn{3}{c}{Accuracy} & \multicolumn{2}{c}{\#MZI $(\times10^{4})$} & \multirow{2}*{\#MZI Red.}
        \\ \cmidrule(lr){2-4} \cmidrule(lr){5-6}
         & Orig. & RVNN & Prop. & Orig. & Prop. & \\
        \midrule
        FCNN        & 98.35\%   & 98.32\%   & 98.02\%   & 31.7  & 7.9     & 75.03\%   
        \\ 
        LeNet-5       & 79.24\%   & 76.57\%   & 75.65\%   & 11.5  & 2.9     & 74.62\%
        \\ 
        ResNet-20   & 93.39\%   & 92.85\%   & 91.10\%   & 116.6 & 29.1    & 75.06\%
        \\ 
        ResNet-32   & 71.89\%   & 69.51\%   & 69.12\%   & 205.1 & 51.5    & 74.88\%
        \\
        \bottomrule
    \end{tabular}
    \label{table:result_overview}
    \end{center}
    \vspace{-0.5cm}
\end{table}

Table \ref{table:result_overview} summarizes the results showing the area efficiency and performance of our proposed work. We compared the proposed OplixNet (Prop.) with the original ONN (Orig.) proposed in \cite{ONN_the_first} in terms of accuracy and area usage. In terms of the real-to-complex data assignment in OplixNet, We apply ``spatial interlace'' to the FCNN model and ``channel lossless'' to the other three CNN models, as is discussed in Section \ref{subsec: optical input encoder}. Due to various computation methods for the optical network area, we utilize the number of MZIs rather than the actual physical area to denote the spatial efficiency of different network architectures. In addition, the column RVNN shows the accuracy of the corresponding software-trained RVNN as a reference. RVNN is the software real-valued version of the given model. Compared with the original ONN, our OplixNet achieves around 75\% area reduction with 0.33\%-3.59\% accuracy degradation. 
In addition, the accuracy difference between OplixNet and RVNN is only 0.30\%-1.48\%.

\begin{figure}[t]
    \centering
    \includegraphics[scale=0.43]{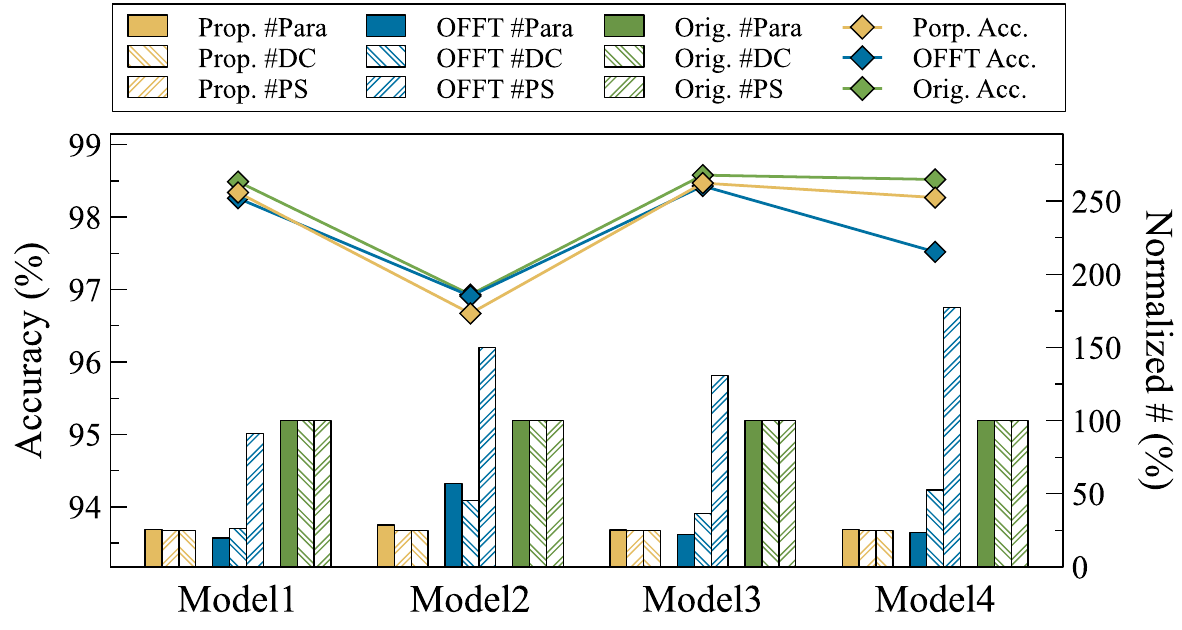}
    \vspace{-0.25cm}
    \caption{Comparison with previous ONN structure \cite{onn_OFFT} under four different FCNN models: Model1-(28$\times$28)-400-10, Model2-(14$\times$14)-70-10, Model3-(28$\times$28)-400-128-10 and Model4-(14$\times$14)-160-160-10.}
    \vspace{-0.4cm}
    \label{fig:compare_gu}
\end{figure}

To demonstrate the performance and efficiency of the OplixNet framework, we conducted a comparison experiment with the previous OFFT structure \cite{onn_OFFT} on different FCNNs. We compared the inference accuracy, number of parameters, DCs, and PSs. Note that all values, such as \#para, \#DC and \#PS, are normalized to the original ONN's corresponding values. For a fair comparison, we use the same hyper-parameter as \cite{onn_OFFT}. In addition, we use the same MZI structure, which contains 2 DCs and 1 PS. As shown in Figure \ref{fig:compare_gu}, the experiment contains four different models. For Model1, its original structure is a 2-layer network, with 784 (28$\times$28) input channels, 400 output channels in the first layer, and 10 output channels in the last layer. Therefore, we denote the configuration of Model1 as (28$\times$28)-400-10. Similarly, the configurations of Model2, Model3 and Model4 are (14$\times$14)-70-10, (28$\times$28)-400-128-10 and (14$\times$14)-160-160-10, respectively. For OFFT and our proposed OplixNet, we use the corresponding configurations to their original version. For Model1, Model3 and Model4, OplixNet outperforms OFFT in inference accuracy. The \#Para in the experiment means the weight parameter number before being mapped into the optical circuit. 
Although OplixNet has more weight parameters in Model1, Model3 and Model4, it has 
fewer DCs and PSs than OFFT after mapping into optical circuits, which means more area is saved in OplixNet.

\begin{figure}[t]
    \centering
    \includegraphics[scale=0.33]{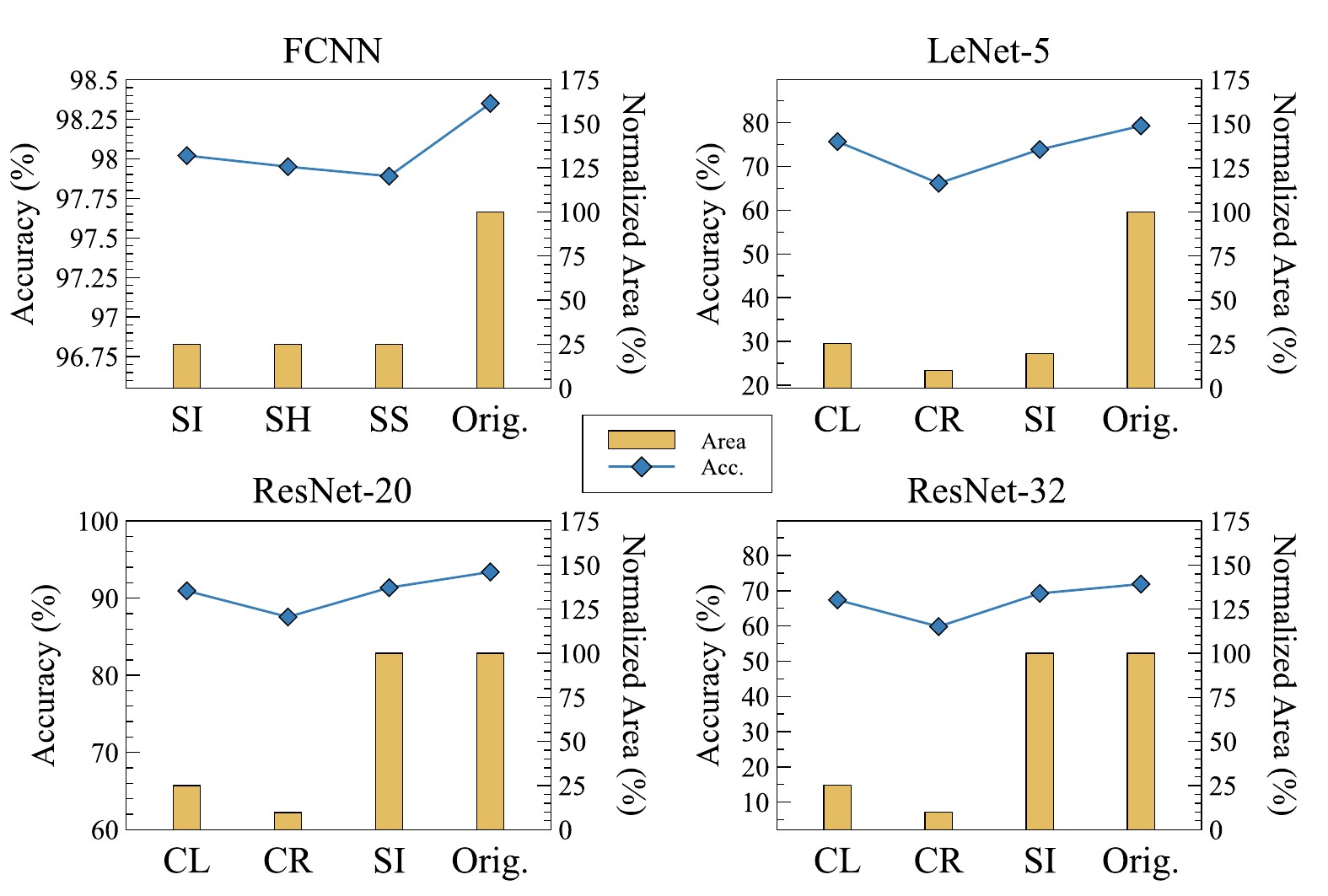}
    \vspace{-0.3cm}
    \caption{Comparison of data assignment on FCNN and CNNs (SI - Spatial Interlace; SH - Spatial Half-half; SS - Spatial symmetric; CL - Channel Lossless; CR - Channel remapping)}
    \vspace{-0.6cm}
    \label{fig:exp2}
\end{figure}

To evaluate the influence of different real-to-complex data assignment, we tested different data assigning methods on four models, as shown in Figure \ref{fig:exp2}. For FCNN, as there is only one channel in MNIST, we test different spatial assigning methods, including ``spatial interlace'' (SI), ``spatial half-half'' (SH) and ``spatial symmetric'' (SS). These spatial assigning methods have the same area reduction of 75.03\%. The SI, which is used in our work for FCNN, has the highest accuracy. This means assigning two neighboring pixels rather than two pixels far away into one complex point is the most recommended way. For the other three CNN models, both spatial and channel assignments are tested, including channel lossless (CL), channel remapping (CR) and the aforementioned SI. The mapping function of CR is from \cite{2D_color_map}.
As is discussed in Section \ref{subsec: optical input encoder}, the spatial assigning methods cannot reduce the area of CONV layers. In LeNet -5, the area reduction of SI is due to the decrease of feature map size in the last few linear layers after flattening. Although SI in LeNet-5 has a slightly larger area reduction (5.8\%), its accuracy degradation is also larger, which is 5.4\% to the original one, while the CL is 3.59\%. A similar problem also exists in CR method, which achieves around 90\% area reduction, causing 5.83\%$\sim$13.12\% accuracy degradation. Note that in ResNet models, there is no area reduction for SI, because the feature map size in the linear layers of ResNet only depends on the channel number of the input data rather than the spatial size. Therefore, the assigning method CL used in OplixNet obtained the best trade-off between area and accuracy for all CNN models.

\begin{table}[t]
    \caption{Results of SCVNN-CVNN Mutual Learning}
    \vspace{-0.45cm}
    \begin{center}
    \begin{tabular}{*{14}{c}}
        \toprule
        \multirow{2}*{Model} & \multirow{2}*{Dataset} & \multirow{2}*{Acc. w/o ML} & \multicolumn{2}{c}{ML with CVNN}
        \\ \cmidrule(lr){4-5}
        & & & Acc. w/ ML & teacher
        \\\midrule
        LeNet-5      & CIFAR10   & 75.65\%   & 75.61\%   & LeNet-5
        \\
        ResNet-20    & CIFAR10   & 90.95\%   & 91.37\%   & ResNet-56
        \\
        ResNet-32    & CIFAR100  & 67.41\%   & 69.12\%   & ResNet-56
        \\
        \bottomrule
    \end{tabular}
    \vspace{-0.6cm}
    \label{table:kd}
    \end{center}
\end{table}

To demonstrate the improvement in accuracy of complex knowledge transfer, we conduct an experiment on three CNN models. In each experimental configuration, we chose the large model in the same series as the teacher model, i.e. ResNet-56 for ResNet-20 and ResNet32. 
For LeNet-5, another LeNet-5 was used as its teacher. 
Additionally, the teacher model adopt CVNN structure, which is the same optical structure as \cite{ONN_the_first}. We use the same mutual learning strategy as \cite{mutual_learning}. The mixing factor $\alpha$ of distillation was set to 1.0 and other training hyper-parameters were kept the same as previous experiments. The experimental results are shown in Table \ref{table:kd}. According to the comparison between with and without mutual learning, large networks like ResNet-32 can get higher accuracy increase, which is 1.71\%. This is because large model has higher redundancy and, therefore, has more representational capacity to learn from their CVNN teacher.

\begin{figure}[t]
    \centering
    \vspace{-0.2cm}
    \includegraphics[scale=0.31]{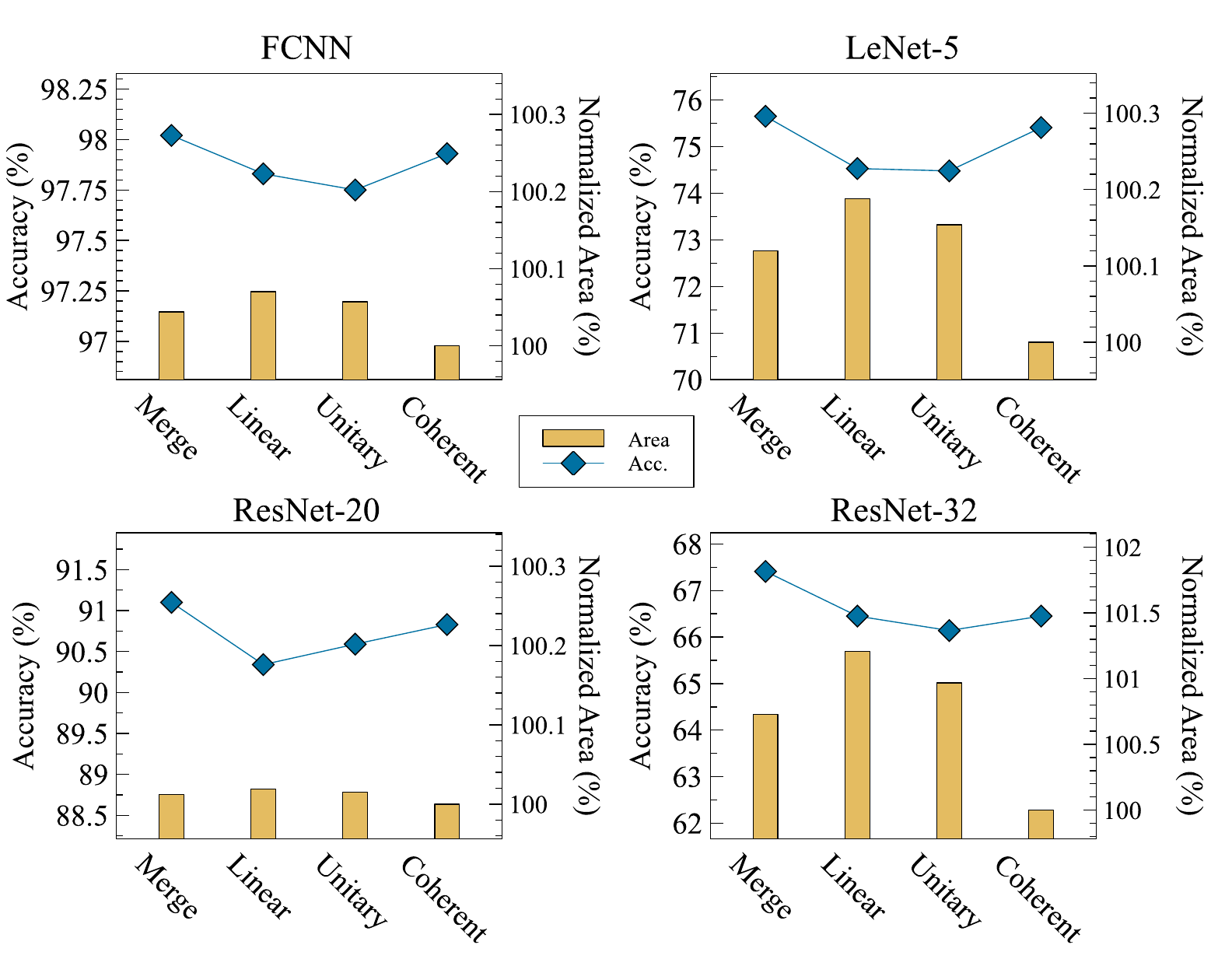}
    \vspace{-0.5cm}
    \caption{Comparison of different decoder settings under different models}
    \vspace{-0.6cm}
    \label{fig:exp_decoders}
\end{figure}

To demonstrate the effectiveness of our learnable output decoder,  an experiment on different output approaches was conducted. As is shown in Figure \ref{fig:exp_decoders}, our ``Merge" decoder was compared with the ``Linear" decoder, the ``Unitary" decoder, and the ``Coherent" approach proposed in \cite{complex_onn}. The area of different configurations is normalized, making the ``Coherent" area become $100\%$ in each model. Note that the approach in \cite{complex_onn} needs additional reference signal, phase shifting time and related post-processing work. Compared with the other two learnable decoders, our ``Merge" decoder has higher accuracy and lower area usage in four models. Compared with \cite{complex_onn}, our ``Merge" decoder achieves 0.09\%$\sim$0.96\% higher accuracy, no shifting time and no further processing work, with a slightly 0.04\%$\sim$0.73\% more area occupation, which is acceptable.

\section{Conclusion}
\label{sec:conclusion}

In this paper, an area-efficient split-complex optical neural network framework, OplixNet, is proposed. Our OplixNet makes full use of the complex property of ONN. Our proposed work consists of 4 parts, including an optical input encoder, input data assigning method, SCVNN-CVNN mutual learning, and design of a learnable output complex decoder. With 0.33\%$\sim$3.59\% accuracy degradation, our proposed framework accomplishes 74.62\%$\sim$75.06\% area reduction.

\section*{Acknowledgement}
\small
This work is supported by TUM International Graduate School of Science and Engineering (IGSSE), Deutsche Forschungsgemeinschaft (DFG, German Research Foundation) – Project-ID 497488621, NSFC (Grant No. 62034007, 62141404), the Major Program of Zhejiang Provincial NSF D24F040002, and SGC Cooperation Project (Grant No. M-0612).

\let\oldbibliography\thebibliography
\renewcommand{\thebibliography}[1]{%
\oldbibliography{#1}%
\fontsize{6.2pt}{6.2}\selectfont
\setlength{\itemsep}{0.03pt}%
}

\bibliographystyle{IEEEtran}
\bibliography{ref}

\end{document}